\documentclass[a4paper,fleqn]{cas-dc}

\usepackage[numbers]{natbib}

\def\tsc#1{\csdef{#1}{\textsc{\lowercase{#1}}\xspace}}
\tsc{WGM}
\tsc{QE}
\tsc{EP}
\tsc{PMS}
\tsc{BEC}
\tsc{DE}

\begin{document}
\let\WriteBookmarks\relax
\def\floatpagepagefraction{1}
\def\textpagefraction{.001}
\shorttitle{Hematite cube swarming}
\shortauthors{O.Petrichenko et~al.}

\title [mode = title]{Swarming of micron-sized hematite cubes in a rotating magnetic field - Experiments}                      



\author[1]{Oksana Petrichenko}[type=editor]
\cormark[1]
\ead{oksana.petricenko@lu.lv}


\address[1]{MMML lab, University of Latvia, Jelgavas 3, Riga, LV-1004, Latvia}

\author[1]{Guntars Kitenbergs}
\author[1]{Martins Brics}

\author[2]{Emmanuelle Dubois}
\address[2]{PHENIX lab, Sorbonne University, CNRS, 4 place Jussieu, Paris, 75252, France}

\author[2]{R\'{e}gine Perzynski}

\author[1]{Andrejs C\={e}bers}


\cortext[cor1]{Corresponding author, E-mail:oksana.petricenko@lu.lv}


\begin{abstract}
Energy input by under-field rotation of particles drives the systems to emergent non-equilibrium states. Here we investigate the suspension of rotating magnetic cubes. Micron-sized hematite cubes are synthesized and observed microscopically. When exposed to a rotating magnetic field, they form rotating swarms that interact with each other like liquid droplets. We describe the swarming behaviour and its limits and characterize swarm size and angular velocity dependence on magnetic field strength and frequency. A quantitative agreement with a theoretical model is found for the angular velocity of swarms as a function of field frequency. It is interesting to note that hematite particles with peanut or ellipsoidal shapes do not form swarms.
\end{abstract}



\begin{keywords}
swarms \sep hematite cubes \sep magnetic particles \sep rotating field 
\end{keywords}

\maketitle

\section{Introduction}
Ensembles of rotating particles, -- so called spinners, recently have obtained growing interest as a particular case of the active matter \cite{1}. 
The benchmark case of these systems is ensembles of magnetic particles, whose rotation may be caused by a rotating field \cite{2,3} or arise by breaking the chirality, for example in the case of phototactic hematite particles \cite{4}. 
Rotating crystals of biological origin were observed in \cite{5}, where fast moving rotating bacteria accumulated on the cell walls due to the hydrodynamic interaction formed collectively rotating ensemble. 
Recently a model based on the account for the lubrication forces between close rotating particles has been introduced \cite{PRE2019}. 
It predict characteristic features of an aggregate of rotating particles -- edge effect in ensembles of rotating particles, stick-slip motion, rationalized by the Frenkel--Kontorova like model.

The article begins with a materials and methods section, which includes the description of hematite cube synthesis, size and magnetic property characterization, microscopy observation methods and subsequent image processing tools.
Next section depicts observation of swarm formation in a qualitative manner.
Further we focus on characterizing swarm angular velocity both qualitatively and quantitatively.
In section \ref{sec:results} we use a a theoretical model from \citet{PRE2019} to explain the peculiarities of hematite particles swarming in the rotating field by comparing it to experimental data.
We finish by discussing the differences and drawing conclusions.

\section{Materials and methods}
\subsection{Synthesis}
The synthesis of hematite cubic particles is based on the gel-sol method of Sugimoto \textit{et~al.} \cite{Sugimoto94,Sugimoto93}.

Sodium hydroxide solution in water ($5.4$~M) is gradually added to iron chloride solution of water ($2.0$~M). 
During the mixing process, solutions are stirred and the temperature increased till $75^{\circ}$C. 
It is important to mark that the salt of the iron chloride should be crystallohydrate ($\rm FeCl_{3}\cdot 6 H_{2}O$). 
Finally, the mixture is hermetically sealed and left in a oven at 100$^{\circ}$C for 7 days.
The formation of cubic (or pseudo-cubic) hematite proceeds by transformations of the iron oxide ($\rm Fe(OH)_{3}$) through akaganeite ($\beta$-$\rm FeOOH$) to hematite.
In such a way precursor, which includes both micron sized particles and nanoparticles of hematite, is obtained \cite{Sugimoto93}.
Washing (performed with ultrasonication and centrifugation) is the last step to obtain the micron sized hematite particles, which are of interest in this study. 
Finally, hematite particles are functionalized with sodium dodecylsulfate (SDS).
The pH of the resulting solution is adjusted by tetramethylamonium hydroxide solution to $8.5-9.5$ values. This procedure avoids hematite particles from irreversible sticking on the glass surface, induced by attractive van der Vaals interactions \cite{Massana_Cid_2017}.
\subsection{Characterization}
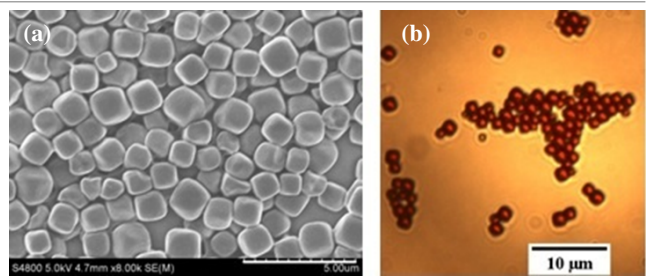
\begin{figure}
	\centering
		\includegraphics[width=\columnwidth]{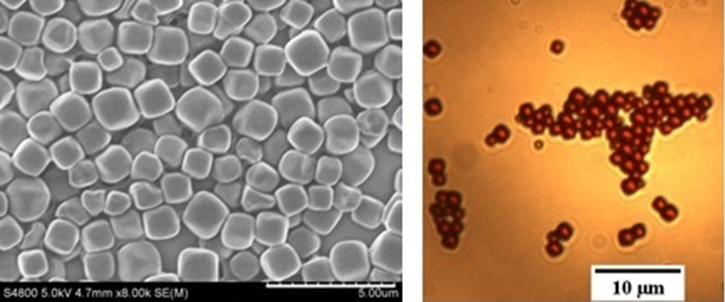}
\begin{picture}(0,0)
	\unitlength=\textwidth
	\textcolor{white}{\put(-0.23,0.2){\bf{(a)}}}
	\textcolor{white}{\put(0.05,0.2){\bf{(b)}}}
	\end{picture}
	\caption{Images of hematite particles obtained by (a) scanning electron microscopy and (b) optical microscopy. Cubic-shape particles have a mean edge length $a\approx1.6\pm 0.3$~$\mu$m.}
	\label{FIG:Cubes}
\end{figure}
The synthesized cubic-shape hematite particles are characterized with microscopy techniques to investigate their size.
For that hematite particles are magnetically stirred in a freshly prepared solution of SDS in water, followed by pH adjustment ($8.5-9.5$). 

Electron microscopy measurements are done with a dried sample of hematite suspension, using a scanning electron microscope Hitachi S4800 at $5.0$ kV and $2000\times$ magnification.
While, for optical microscopy a capillary of $100$~$\mu$m thickness and $2$~mm width (VitroCom Vitrotubes\texttrademark) is then filled with the suspension of cubes via capillary forces and sealed with paraffin.
It is imaged with $100\times$ magnification.
As seen in images of Fig.\ref{FIG:Cubes}, obtained by both techniques, the resulting particles have a cubic shape with a similar edge length $a = 1.6\pm 0.3$~$\mu$m.
In addition, while observed with an optical microscope, cubes tend to form fluctuating aggregates, where attraction is due to magnetic and van der Vaals interactions and fluctuations come from Brownian motion.

Magnetic properties for a dried sample are determined by a vibrating sample magnetometer (Lake Shore Cryotronics, Inc., 7400 VSM).
The magnetization curve (not shown here) indicates a clear hysteresis and notable coercivity ($H_c=3.5~\text{kG}$).
Remanent magnetization is around $\sigma=0.2~\text{emu/g}$, which corresponds to a particle magnetization $M=1.1~\text{G}$.

\subsection{Video microscopy}
Experiments were done in two laboratories in Riga and Paris.
System in Riga is based on an inverted microscope Leica DMI3000B.
A custom made coil device with 2 perpendicular pairs of coils provide homogeneous field in the plane of observation.
Coils are run with {Kepco~BOP~20-10M} power supplies, which are controlled with a NI DAQ card via a LabView program.
Sinusoidal signals with a $90^{\circ}$ phase difference creates a homogeneous rotating field at the field of observation up to $1$~kHz.
Process is filmed in brightfield mode with a color camera Leica DFC310 FX ($1.4$MP).
Using $100\times$ objective with oil immersion and $0.70\times$ camera mount provides a $130\times95$~$\mu\text{m}^2$ field of view.

System in Paris is very similar and is described in \cite{JFMdrops}.
In comparison, it uses $40\times$ objective, but allows to make recordings with a synchronized magnetic field up to $60$~Oe and a frame rate up to $60$~fps.

\subsection{Image analysis}
\label{sec:imageanalysis}
Experimental images are analyzed both qualitatively and quantitatively. For qualitative analysis, ImageJ program is used, to analyze individual cubes, observe aggregates and swarm formation and approximate sizes.  

For quantitative analysis, a MatLab code for swarm angular velocity and size determination is developed.
First, to find the an aggregate or swarm, each image is inverted and thresholded.
Using the intensity information only coming from pixels of particle aggregates allows to remove background.
Then the center of aggregate is found by calculating the center of mass.
Aggregate size is found by averaging intensities $I$ in polar coordinates, which are normalized by total aggregate intensity.
For aggregate radius $R$ we take the size at initial state.
Swarm angular velocity is found by cross-correlation of two images.
To exclude magnetic field direction influence (swarms tend to consist of short chains of cubes that rotate with the field), images with the same field direction are used.
Second image is rotated around the centre of mass by multiple angles and cross-correlated with the first image.
Both images are normalized, therefore correlation peak value can be used to identify the angle between swarms in the two images. 
Dividing it with the time between images, one finds swarm angular velocity $\Omega$.
It is averaged over the set of images for each frequency, obtaining the mean value and its error.

\section{Results and discussion}
\label{sec:results}

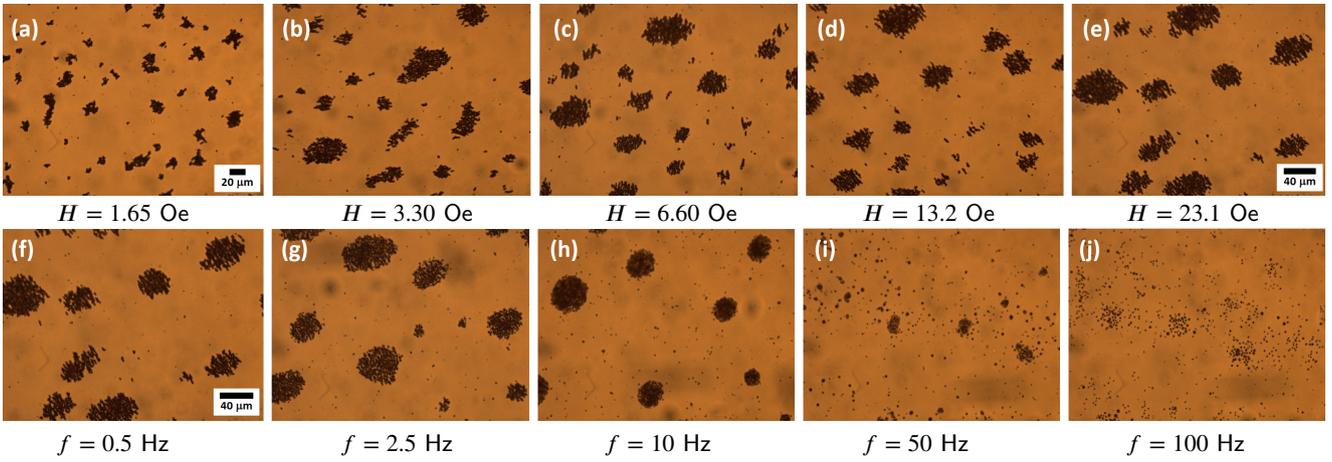
\begin{figure*}
	\centering
	\includegraphics[width=\textwidth]{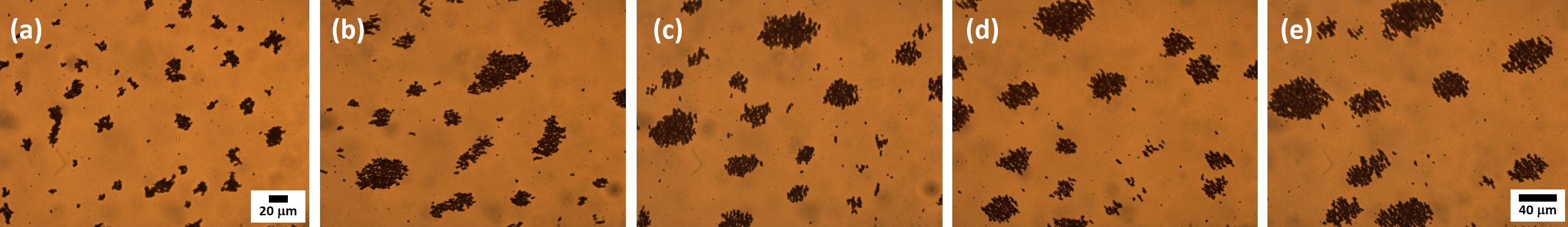}
		\begin{picture}(0,0)
	\unitlength=\textwidth
	{\put(-0.46,0.003){$H=1.65~\text{Oe}$}}
	{\put(-0.245,0.003){$H=3.30~\text{Oe}$}}	{\put(-0.045,0.003){$H=6.60~\text{Oe}$}}
	{\put(0.15,0.003){$H=13.2~\text{Oe}$}}
	{\put(0.35,0.003){$H=23.1~\text{Oe}$}}
	\end{picture}
	\includegraphics[width=\textwidth]{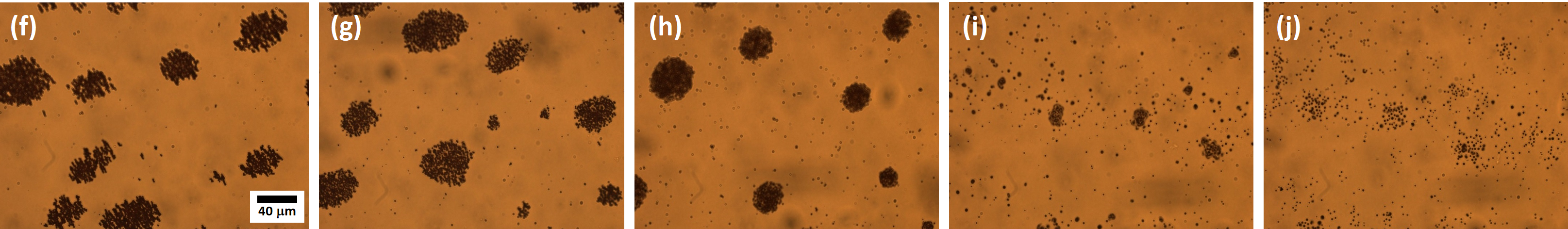}
		\begin{picture}(0,0)
	\unitlength=\textwidth
	{\put(-0.46,-0.001){$f=0.5~\text{Hz}$}}
	{\put(-0.245,-0.001){$f=2.5~\text{Hz}$}}
	{\put(-0.045,-0.001){$f=10~\text{Hz}$}}
	{\put(0.15,-0.001){$f=50~\text{Hz}$}}
	{\put(0.35,-0.001){$f=100~\text{Hz}$}}
	\end{picture}

	\caption{Swarming of hematite cubes under rotating magnetic field. First row (a)--(e) shows swarm development with an increasing magnetic field strength at a constant frequency $f=0.5~\text{Hz}$. 
	Second row (f)--(j) shows change of swarm behavior with an increasing frequency at a constant field  $H=23.1~\text{Oe}$.}
	\label{FIG:Swarms}
\end{figure*}
\begin{figure}
	\centering
	\includegraphics[width=0.8\columnwidth]{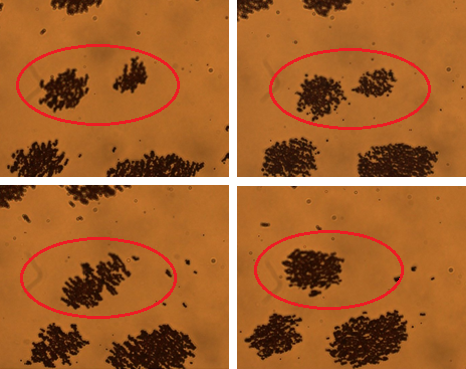}
	\begin{picture}(0,0)
	\unitlength=\textwidth
	\textcolor{white}{\put(-0.232,0.28){(a)}}
	\textcolor{white}{\put(-0.045,0.28){(b)}}
	\textcolor{white}{\put(-0.245,0.13){(c)}}
	\textcolor{white}{\put(-0.06,0.13){(d)}}
	\end{picture}
	\caption{Fusion of two swarms at $H\approx20~\text{Oe}$ and $f\approx0.6~\text{Hz}$. (a)~$t=0~\text{s}$, (b)~$t=164~\text{s}$ ($2.7$~min), (c)~$t=192~\text{s}$ ($3.2$~min), (d)~$t=295~\text{s}$ ($4.9$~min)}
	\label{FIG:Fusion}
\end{figure}

When a suspension of hematite cubes is exposed to an in-plane rotating magnetic field, we observe a spontaneous swarming (see Fig.\ref{FIG:Swarms}).
We start our experiment with a low and constant field frequency ($f=0.5~\text{Hz}$), while gradually increasing the magnetic field strength.
At $H = 1.65~\text{Oe}$ (Fig.\ref{FIG:Swarms}(a)) hematite particles form single and double chains, that rotate around their axes and form swarm-like aggregates. 
An increase of field ($H = 3.30-9.90~\text{Oe}$, (Fig.\ref{FIG:Swarms}(b)--(c)) enhances swarming and we observe bigger elongated swarms, which consist of multiple rotating chains. 
Longer semi-axis of observed swarms reach $75~\mu\text{m}$.
By further increase of field $H = 13.2-23.1~\text{Oe}$ (Fig.\ref{FIG:Swarms}(d)--(e)) swarm formation continues, however, their further increase in size is limited by the number of surrounding particles.
Cohesion forces responsible for swarm formation imply that swarms behave as liquid droplets.
When they come close enough, swarms merge into one, as shown in Fig.\ref{FIG:Fusion}.

We then continue the experiment with increasing the frequency, while magnetic field remains constant at $H$=23.1~\text{Oe}.
Characteristic images are given in Fig.\ref{FIG:Swarms}(f)--(j).
Increasing the frequency up to $2.5~\text{Hz}$ induce a structural change in the aggregate--swarm behavior.
Previously observed chains have split in individual cubes and dimers and the swarm shape becomes more regular (Fig.\ref{FIG:Swarms}(g)).
Further increase of frequency to $f=10~\text{Hz}$ transforms swarms to more compact and circular shapes.
Some individual cubes escape the swarms, but also return to them, if they come close enough (Fig.\ref{FIG:Swarms}(h)).
Increasing frequency even more ($f\geq50~\text{Hz}$) induces disintegration of swarms, as can be seen in Fig.\ref{FIG:Swarms}(i--j).

It is interesting to note that hematite particles with peanuts or ellipsoid shapes do not form swarms.

\begin{figure}
\centering
\includegraphics{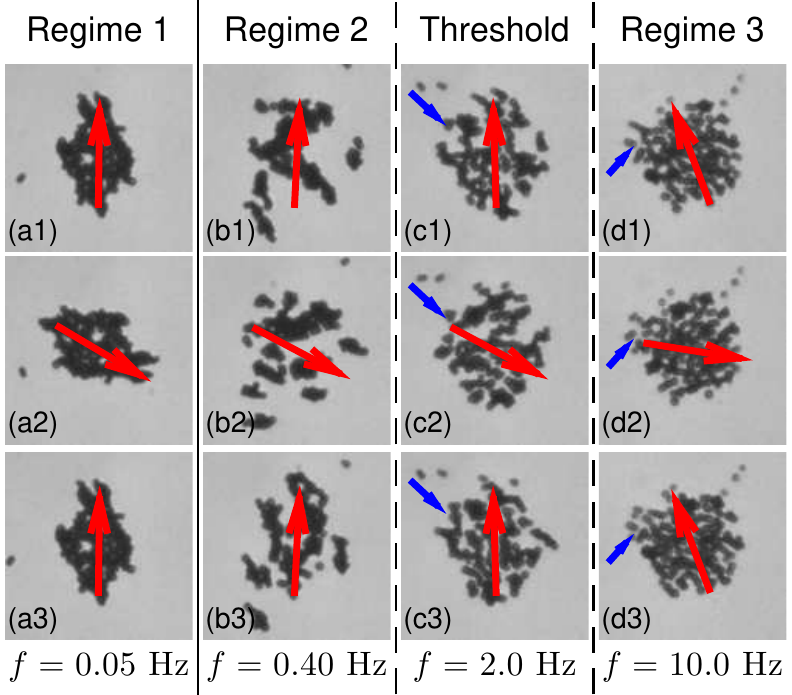}
\caption{Aggregate rotation dynamics dependence on field frequency can be split in 3 regimes: single aggregate, rotates with the field (a), multiple aggregates recombining, rotating with the field (b) and single cubes and dimers rotating with the field and forming a rotating swarm (d). (c) marks the threshold between regimes 2 and 3. Large red arrow indicates the direction of the field. Small blue arrow marks a reference point to see particle displacement. Images are taken for a swarm with initial radius $R=10~\mu$m at $H=33$~Oe.}
\label{FIG:Rotation}
\end{figure}

As mentioned in the introduction, recent theoretical results on rotating field driven ensembles \cite{PRE2019} motivate us to investigate swarm rotation experimentally in higher detail.
We perform a detailed study of several swarms using the setup in Paris, where magnetic field direction is known for each image.
Two rotating field amplitudes are exploited ($33~\text{Oe}$ and $58~\text{Oe}$), while varying frequencies from $0.01~\text{Hz}$ to $50.0~\text{Hz}$.
Characteristic images can be seen in Fig.\ref{FIG:Rotation}.
They have been selected to show the rotation behavior.
For each frequency series of three images are chosen.
First series corresponds to a magnetic field direction as close as possible to upwards direction.
Second corresponds to a field direction at $120^\circ$ with respect to the first one.
Third corresponds to the image with field direction pointing in the same direction as in the first image.
These conditions correspond to the following times: for (a) $t_\text{(a1)}=0.0$~s, $t_\text{(a2)}=6.7$~s, $t_\text{(a3)}=20.0$~s; for (b) $t_\text{(b1)}=0.0$~s, $t_\text{(b2)}=0.8$~s, $t_\text{(b3)}=10.0$~s; for (c) $t_\text{(c1)}=0.0$~s, $t_\text{(c2)}=0.17$~s, $t_\text{(c3)}=0.5$~s and for (d) $t_\text{(d1)}=0.0$~s, $t_\text{(d2)}=0.03$~s, $t_\text{(d3)}=0.20$~s.
Magnetic field direction is shown with a large red arrow.
For (c)\,\&\,(d) a small blue arrow indicates a point of reference.

\begin{figure}
	\centering
\includegraphics{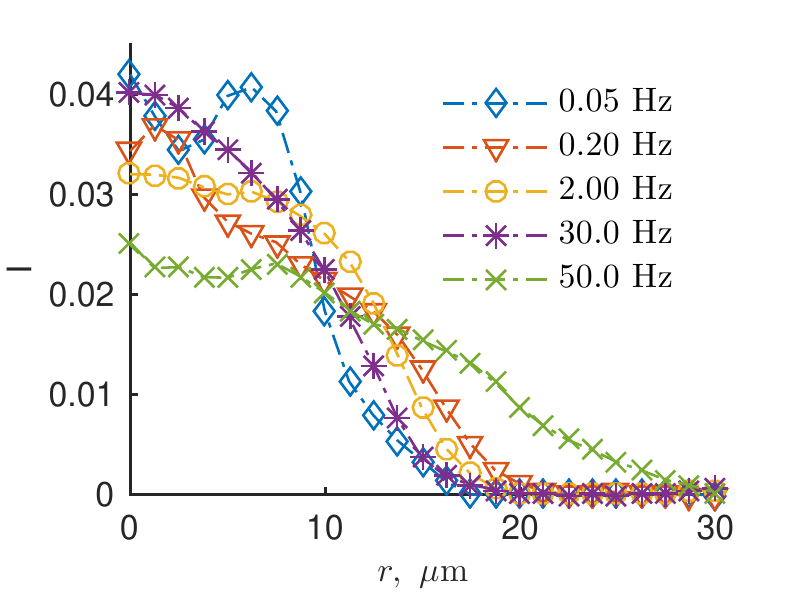}
\caption{Characterization of aggregate size by the distribution of intensity $I$ dependence on radial distance $r$ at various field frequencies $f$. Initial radius $R\approx10~\mu$m, magnetic field is $H=33~\text{Oe}$.}
\label{FIG:SwarmSize}
\end{figure}

Rotational behavior of aggregates--swarms as a function of frequency $f$ at a given field can be split in 3 regimes (see Fig.\ref{FIG:Rotation}).
This behaviour also reflects in aggregate--swarm size, which is characterized in Fig.\ref{FIG:SwarmSize}.

Regime 1: If the magnetic field frequency is very low ($f<0.1$~Hz), cubes form an aggregate that rotates as a solid with the field.
For example, see Fig.\ref{FIG:Rotation}(a) and a supplementary video S1.
As the cubes are almost close packed in the aggregate, the radius is smallest (blue diamonds in Fig.\ref{FIG:SwarmSize}, $R\approx10~\mu\text{m}$).

Regime 2: Increase of frequency splits the large aggregate in several smaller aggregates that collide with each other and rotate with the field.
An example can be seen in Fig.\ref{FIG:Rotation}(b) and a supplementary video S2.
The transition to this second regime happens already around $f$=\,0.10~\text{Hz}.
Initially this spreads the aggregate and slightly increases its size (red triangles in Fig.\ref{FIG:SwarmSize}).
A further frequency increase reduces the size of hematite aggregates in the aggregate, which rotate closer to each other, slightly decreasing the aggregate size (orange circles in Fig.\ref{FIG:SwarmSize}), until at around $f=2.0~\text{Hz}$ only single cubes or dimers can be observed.
This regime contains the transition from a single aggregate (Regime 1) to a swarm (Regime 3).

Regime 3: Swarms (aggregates that consist of rotating single cubes or dimers) have a clearly circular shape, as can be seen in Fig.\ref{FIG:Rotation}(c)\,\&\,(d) and a supplementary video S3.
In addition, increase of the field frequency enables faster rotation of the swarm.
This continues to reduce the size (violet asterisks in Fig.\ref{FIG:SwarmSize}), almost to the average size of the large aggregate in Regime 1.

If field frequency is increased even further ($f\approx50~\text{Hz}$), the swarms disassemble.
This is clearly visible by the size increase (green crosses in Fig.\ref{FIG:SwarmSize}).

Further we analyze swarm rotation quantitatively via the method described in section \ref{sec:imageanalysis}.
Characterization of swarm rotation is only reasonable for the third regime.
In the first regime ($f<0.2~\text{Hz}$) particle aggregate rotates as a whole (solid rotation) and does not behave like a swarm of multiple particles.
In the second regime ($0.2<f<2.0~\text{Hz}$) several smaller rotating aggregates that collide and reassemble in various combinations, making it impossible to define a collective angular velocity.

\begin{figure*}
	\centering
\includegraphics{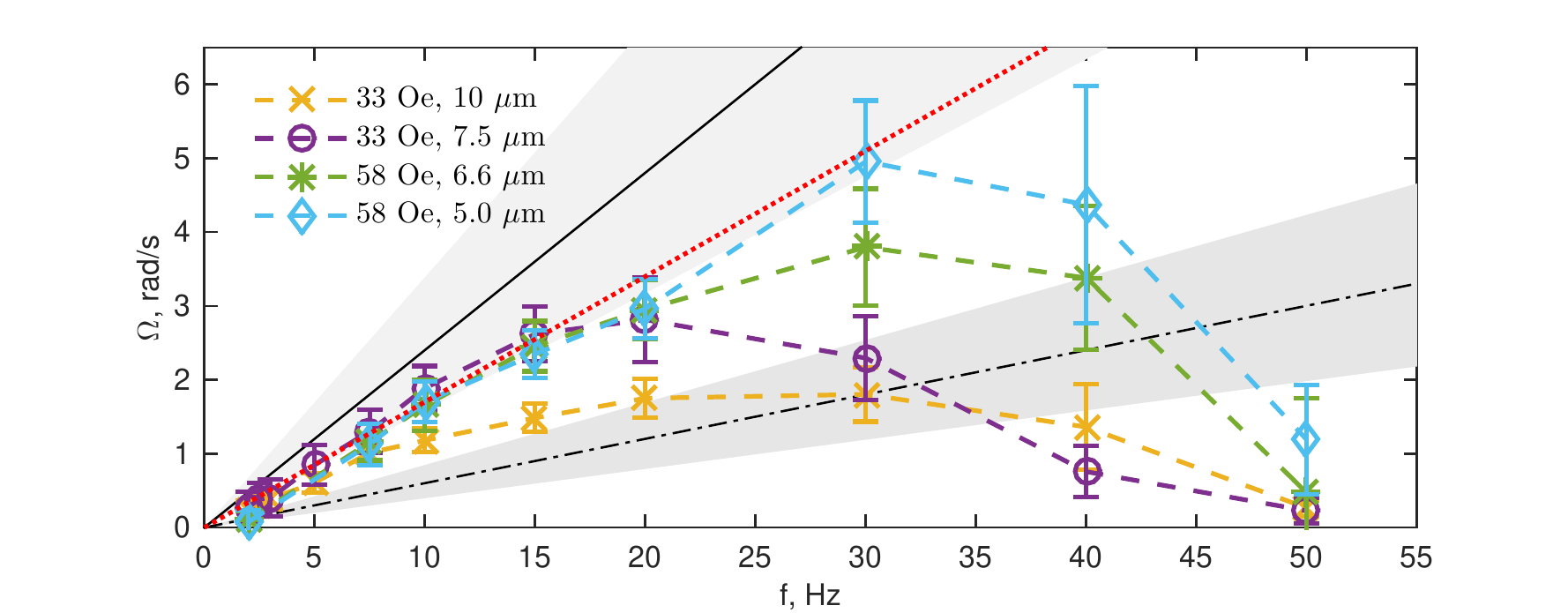}
\caption{Swarm angular velocity $\Omega$ as a function of field frequency $f$ for four swarms of different sizes and magnetic fields, as indicated in the legend. Red dotted line shows model prediction for $R/d=7.0/1.6$. black dash-dotted line for $R/d=10/1.6$, solid black line for $R/d=5/1.6$. Gray areas indicate corresponding uncertainties to black lines, which come from particle size distribution.}
\label{FIG:RotGraph}
\end{figure*}

For the third regime many individual particle rotation induces a gloal rotation of the swarm.
The swarm angular velocity $\Omega$ is presented in Fig.\ref{FIG:RotGraph} as a function of the field frequency $f$ for swarms of different radii $R$ at two different field strengths $H$.
For frequencies larger than $f=2.0$~Hz, swarm angular velocity first increases linearly.
It is important to note that at this stage the swarm rotation frequency is $\approx30$ times smaller that the magnetic field frequency.
When a critical frequency $f_c=20..30$~Hz is reached, the swarm rotation slows down.
At $f=40..50$~Hz angular velocity $\Omega$ goes quickly to zero, as swarm starts to disassemble and individual particles move far from each other.

Increase in the magnetic field strength $H$ increases the critical frequency $f_c$ and allows to reach higher swarm angular velocities (green asterisks and blue diamonds in Fig.\ref{FIG:RotGraph}), while swarms with a larger initial radius $R$ rotate slower (orange crosses vs. violet circles in Fig.\ref{FIG:RotGraph}). To have a more quantitative view, further measurements are necessary.

Data obtained for the angular velocity of swarm $\Omega$ as a function of the magnetic field frequency $f$ (particle rotation frequency) reasonably well agree with the relation derived from the balance of lubrication forces at the edge of swarm and friction near a solid wall in \cite{PRE2019}
\begin{equation}
\frac{\tau\Omega}{F_{d}}=\frac{\pi}{\varphi_{d}}\frac{1}{(R/d)^2} ,
\end{equation}
\noindent
where \,\,$\tau=d/(3m^{2}/2d^{4}\zeta)$\quad is a characteristic timescale, ${F_{d}=\dfrac{1}{5}\ln{(d/\delta)}\dfrac{2\pi\eta 2\pi f d^{6}}{3m^{2}}}$ is the driving force,
$d$ is is the size of particle, here assimilated to cube edge size $a$, $m$ is its magnetic moment, $\delta$ is the distance between surfaces of particles and $\varphi_d$ is the area fraction of particles. 
We can estimate hydrodynamic drag coefficient near solid wall as $\zeta=3\pi\eta d \ln{(d/\delta)}$ and instead of $\pi/\varphi_{d}$ take a value $5.6$, which is more precise and obtained numerically in \cite{PRE2019}.
As a result, we get
\begin{equation}
\frac{\Omega}{f}=\frac{11.2\pi}{15}\frac{1}{(R/d)^{2}}.
\end{equation}
For a swarm with a radius $R=7.0~\mu$m, taking particle size $d=a=1.6~\mu$m, this gives $\Omega/f=0.17$, which is very close to the experimental data, as indicated by red dotted line in Fig.\ref{FIG:RotGraph}.
Dash dotted black line shows model prediction for a swarm of $R=10~\mu$m, while solid black line shows prediction for a swarm of $R=5~\mu$m.
Using $d=a=1.6~\mu$m, the uncertainty of the measurement $\pm0.3~\mu$m affects the model prediction considerably.
This is displayed in Fig.\ref{FIG:RotGraph} via gray areas, which correspond to the black lines.
Nevertheless, we can see that the linear part of our measurements fits well within these limits.

\section{Conclusions}
Swarming of hematite cubes in the rotating field is investigated experimentally.
It is shown that in the intermediate range of frequency of the rotating field the swarm consists mainly of single rotating particles which drive the rotation of swarms with angular velocities smaller by an order of magnitude of the angular velocity of the field.
The angular velocity of the swarm determined from the cross correlation analysis of images reasonably well corresponds to the theoretical relation with respect to the field frequency and swarm size.
Due to the action of cohesion forces swarms behave like liquid droplets -- two swarms, if close enough, merge similarly to two liquid droplets.
At higher frequencies, when single particles do not follow anymore the rotating field, the swarms break.

\section*{Aknowledgements}
Authors are very thankful to P.Tierno and H.Massana-Cid from the University of Barcelona for the instructions on hematite synthesis. 
Authors also thank colleagues from the University of Latvia -- M.M.Maiorov from the Institute of Physics for magnetic measurements and K.Buks from the Institute of Chemical Physics for SEM images of particles.

O.P. acknowledges support from PostDocLatvia grant No. 1.1.1.2/VIAA/1/16/018, G.K. from PostDocLatvia grant No. 1.1.1.2/VIAA/1/16/197 and M.B. and A.C. from M.era-net project FMF No.1.1.1.5./ERANET/18/04. 
All authors are thankful to French-Latvian bilateral program Osmose project FluMaMi (n$^{\circ}$40033SJ; LV-FR/2019/5). 

\section*{Appendix A. Supplementary videos} 
Supplementary videos associated with this article can be found in the online version at ...

\printcredits

\bibliographystyle{model1-num-names}

\bibliography{cas-refs}

\end{document}